\documentclass[preprint,pre,nofootinbib,showpacs,titlepage,showkeys]{revtex4}%
\usepackage{amsfonts}
\usepackage{amsmath}
\usepackage{amssymb}
\usepackage{graphicx}
\usepackage{acronym}%
\begin{document}
\title{Flow distributed oscillation, \ flow velocity modulation and resonance \ }
\author{Patrick N. McGraw and Michael Menzinger}
\affiliation{Department of Chemistry, University of Toronto, Toronto, Ontario, Canada M5S 3H6}

\begin{abstract}
We examine the effects of a periodically varying flow velocity on the standing
and travelling wave patterns formed by the flow-distributed oscillation (FDO)
mechanism. \ In the kinematic (or diffusionless) limit, the phase fronts
undergo a simple, spatiotemporally periodic longitudinal displacement. \ On
the other hand, when the diffusion is significant, periodic modulation of the
velocity can disrupt the wave pattern, \ giving rise in the downstream region
to travelling waves whose frequency is a rational multiple of the velocity
perturbation frequency. \ We observe frequency locking at ratios of 1:1, 2:1
and 3:1, depending on the amplitude and frequency of the velocity modulation.
\ This phenomenon can be viewed as a novel, rather subtle type of resonant forcing.

\end{abstract}
\pacs{82.40.Ck, 47.70.Fw}
\pacs{82.40.Ck, 47.70.Fw}
\pacs{82.40.Ck, 47.70.Fw}
\pacs{82.40.Ck, 47.70.Fw}
\pacs{82.40.Ck, 47.70.Fw}
\pacs{82.40.Ck, 47.70.Fw}
\pacs{82.40.Ck, 47.70.Fw}
\pacs{82.40.Ck, 47.70.Fw}
\pacs{82.40.Ck, 47.70.Fw}
\keywords{Open reactive flows, reaction-advection-diffusion systems, spatiotemporal resonance}
\maketitle

Interest has been mounting recently on mechanisms of pattern formation in open
reactive flows. \ The combination of reaction, advection and diffusion,
together with the effect of an upstream boundary condition, leads to
mechanisms such as flow-distributed oscillations (FDO) \cite{Kuznetsov}%
-\cite{McGraw}, a general category of stationary patterns referred to as "flow
and diffusion-distributed structures" (FDS) \cite{Satnoianu1}-\cite{McGrawFDS}%
, and the differential flow instability (DIFI) \cite{Rovinsky92}\cite{MRDIFI}.
\cite{Satnoianu1}-\cite{McGrawFDS} \ Our focus here is on FDO. \ Due to the
equivalence \cite{Kaern3}-\cite{Faraday} of flow in
reaction-advection-diffusion (RAD) systems and linear growth of the spatial
domain of a reaction-diffusion system, and the existence of cellular
oscillations in segmenting tissue, FDO was shown to be involved in the axial
segmentation occurring during biological development \cite{Kaern3}%
-\cite{Faraday}. Given the pulsating growth of certain organisms
\cite{Pulsatile1}-\cite{Pulsatile2}, including human embryos \cite{Pulsatile3}%
, we study here the consequences of a periodically modulated flow $v(t)$ on FDO.

The systems of interest are described by the RAD equation without differential
transport:%
\begin{equation}
\frac{\partial\mathbf{U}}{\partial t}=\mathbf{f}(\mathbf{U})-v(t)\frac
{\partial\mathbf{U}}{\partial x}+D\frac{\partial^{2}\mathbf{U}}{\partial
x^{2}}, \label{RDA}%
\end{equation}
where $D$ is the diffusion constant, $v(t)$ is the flow velocity,
$\mathbf{U}(x,t)$\ is an $N$-dimensional vector of dynamical variables
(concentrations of species) and the local dynamics given by the vector valued
rate function $\mathbf{f}(\mathbf{U})$ has an attracting limit cycle. \ If
$v(t)$ is constant, this system can support flow-distributed oscillations
(FDO) controlled by the upstream boundary condition $\mathbf{U}(0,t)$. \ In
the simplest case, \ a constant boundary condition sets the phase of each
oscillating fluid element as it enters the medium, and the periodic recurrence
of the same phase as the fluid travels downstream results in stationary waves.
\ Oscillating boundary conditions result in travelling waves.\cite{Kaern3}%
\cite{Kaern4}\cite{Faraday}\cite{Kaern5} Diffusion can modify the effective
dynamics of the medium as it travels downstream and even extinguish the
oscillations.\cite{Bamforth2}\cite{McGraw} \ Equation (1) is also relevant to
media such as linearly growing organisms, \ as it can be reinterpreted by
means of a Galilean transformation as representing a stationary medium with a
boundary (the growth tip) moving at speed $v(t)$.\cite{Kaern3}\cite{Kaern4}%
\cite{Faraday}\cite{Kaern6} \ 

We examine the effect of a sinusoidally varying velocity $v(t)=v_{0}%
+\delta_{v}\cos\omega_{v}t$. In the \emph{kinematic} limit of vanishing
diffusion $D/v^{2}\rightarrow0$ the wave pattern undergoes a simple,
calculable longitudinal displacement which is periodic in both time and space.
\ \ Away from the kinematic limit, however, we observe a type of nonlinear
resonance. \ Relatively small disturbances of an FDO wave pattern are
magnified with downstream distance until the wavefronts break. \ This rupture
generates travelling waves in the downstream region whose temporal frequency
is a rational multiple of the velocity perturbation frequency. \ We observe
1:1, 2:1 and 3:1 ratios depending on the frequency and amplitude of the
velocity perturbation. \ \ 

For the numerical examples, we use the FitzHugh-Nagumo-type (FN)
dynamics\cite{FNModel}%
\begin{equation}
\mathbf{f}(X,Y)=%
\begin{pmatrix}
\varepsilon(X-X^{3}-Y)\\
-Y+\alpha X+\beta
\end{pmatrix}
\label{FNmodel}%
\end{equation}
with $\varepsilon=5$, $\alpha=2$ and $\beta=0$ for the local rate function.
\ At these parameter values, \ the local system has a limit cycle and a
moderately strong nonlinearity. \ The frequency of the limit cycle oscillation
is $\omega_{0}\approx2\pi(0.43)$. \ In all simulations, we set $D=1$ and vary
only $v_{0}$ and $\delta_{v}$.\ 

When the ratio $D/v^{2}$ is sufficiently small, \ diffusion is relatively
unimportant and each individual fluid element behaves approximately as an
independent oscillator obeying the local dynamics whose initial phase is set
by the boundary condition as it enters the flow from the upstream
end.\cite{Kaern1} \ The phase fronts can then be calculated by pure
kinematics: \ the oscillation phase of a fluid element at a particular time
and location depends on \ its initial phase when it entered the flow and \ how
long ago it entered the flow. \ For the case of a stationary boundary
condition (i.e., constant phase at the boundary) \ the result is that the
location of the phase front for a particular value of the oscillation phase
$\phi$ is given by%
\begin{equation}
x(\phi,t)=v_{0}\frac{\phi}{\omega_{0}}+\frac{\delta_{v}}{\omega_{v}}%
(\sin\omega_{v}t-\sin\omega_{v}(t-\frac{\phi}{\omega_{0}})),
\label{phasefronts}%
\end{equation}
where $\omega_{0}$ is the frequency of the local oscillator. \ \ When
$\delta_{v}=0$, this reduces to the simple linear mapping between position and
phase that characterizes stationary FDO waves. \ \ \ Note that: \ 1) \ \ The
amplitude of the displacement of the phase fronts (second, time-dependent term
in eq. \ref{phasefronts}) \ depends on $\delta_{v}/\omega_{v}$, \ implying
that faster (higher frequency) velocity modulation has less of an effect than
slower modulation. \ \ 2) \ The displacement is periodic in time with the same
frequency as the velocity perturbation, but \ 3) \ it is also periodic in
$\phi$ and thus in space. \ \ The fronts for which $\phi$ is a multiple of
$\pi\omega_{0}/\omega_{v}$ are not displaced, and they occur periodically at
positions \ $\pi nv_{0}/\omega_{v}$. \ \ The spatial periodicity can be
understood by considering the trajectories of fluid elements entering the
system at different times. \ Different elements enter at different points in
the velocity modulation period and thus begin their downstream travel at
different initial velocities. \ Over any multiple of the modulation period,
however, the velocity averages to $v_{0}$ and thus all elements reach the same
position at the same phase when one full period has passed, regardless of when
they started. \ 

When the boundary condition is oscillatory instead of stationary, \ travelling
waves are generated. \ \ Just as in the stationary case, velocity modulation
causes a periodic longitudinal displacement of the travelling wavefronts.
\ \ The modulation of stationary and travelling waves is illustrated in figure
\ref{wigglewaves}. \
\begin{figure}
[ptb]
\begin{center}
\includegraphics[
height=3.4489in,
width=4.1943in
]%
{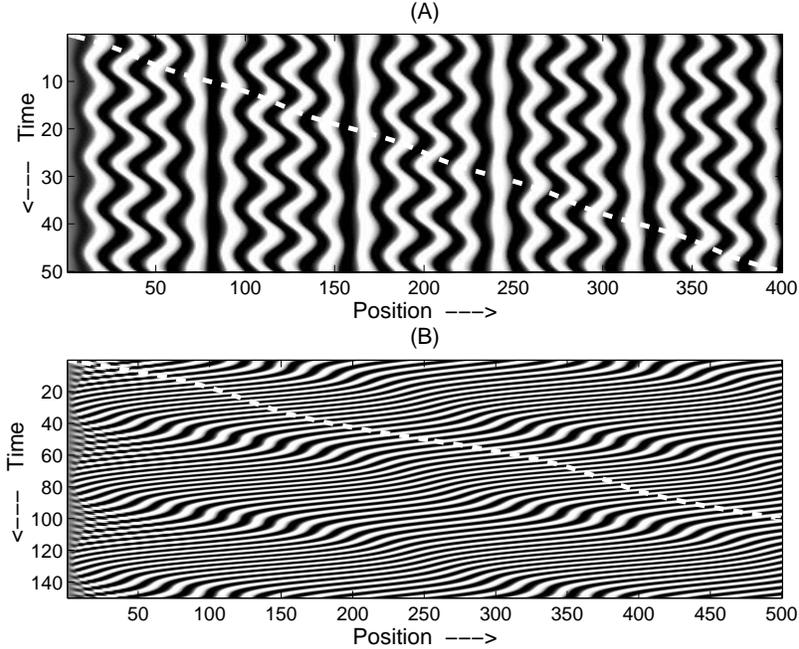}%
\caption{Periodic modulation of standing and travelling FDO waves in the
kinematic limit. \ Note the periodicity in both time and space. \ In each
space-time diagram, the trajectory of a co-moving point is shown as a guide to
the eye (dashed white line). \ (A) Stationary waves: \ $v_{0}=8,$ $\delta
_{v}=2,$ $\omega_{v}=2\pi(0.1)\approx\omega_{0}/4$. \ Some wavefronts remain
stationary while others wiggle back and forth.\ (B) Upstream travelling waves
with wave frequency $\omega_{tw}=2\pi(0.25)$ subject to a modulated velocity
field with $v_{0}=5,$ $\delta_{v}=2,$ $\omega_{v}=2\pi(0.02)$.}%
\label{wiggle-waves}%
\end{center}
\end{figure}

When diffusion is unimportant, neighboring fluid elements do not interact and
the behavior of FDO patterns can be explained by pure kinematics. \ Each
co-moving fluid element follows the limit cycle defined by the batch reactor
dynamics. \ However, significant diffusion alters the dynamics. \ \ The flow
velocity modulation then introduces a periodic variation in the local
environment of each fluid element. \ The strength of the local gradient is
different for each co-moving element and diffusion therefore affects the
dynamics differently at different locations. \ This differential effect can
magnify the small kinematic effect of the velocity perturbation, leading to
larger differences in the dynamical variables and eventually to disruption of
the smooth FDO waveforms. \
\begin{figure}
[ptb]
\begin{center}
\includegraphics[
height=5.5711in,
width=3.4298in
]%
{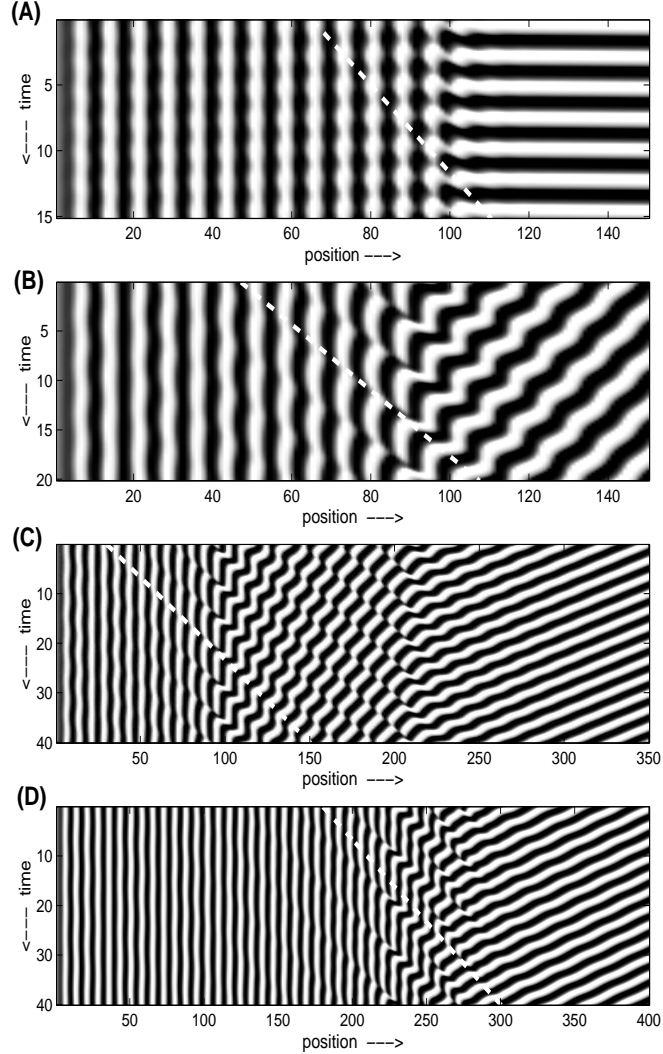}%
\caption{Examples of the breakup of waves due to a velocity perturbation.
\ All examples have average flow velocity $v_{0}=3$.\ \ The dotted white line
in each frame represents $v_{0}$. \ \ A) \ $\omega_{v}=2\pi(0.43)\approx
\omega_{0}$, $\delta_{v}=0.215$. \ A slight pulsation with frequency
$\omega_{v}$ is visible in the standing wavefronts, \ baecoming more
pronounced downstream. \ Near $x=100$ there is a transition to uniform
oscillation. \ \ \ B) \ $\omega_{v}=2\pi(0.2)\approx\omega_{0}/2,$ $\delta
_{v}=0.1$. \ Periodic disturbance of the standing waves becomes sharper with
increasing distance, and there is a transition to travelling waves with a 1:1
frequency ratio. \ Near the transition these travelling waves propagate in a
saltatory manner but they grow smoother with further downstream distance. \ C)
\ $\omega_{v}=2\pi(0.15)\approx\omega_{0}/3$, $\delta_{v}=0.1125$. \ As in (B)
there is a transition to travelling waves near $x=100$, but in this case the
waves do not smooth out with downstream distance. \ Instead, there is a second
transition at x=200 to waves with twice the velocity modulation frequency (2:1
resonance). \ \ D) \ $\omega_{v}=2\pi(0.10)\approx\omega_{0}/4$, $\delta
_{v}=0.05$. A series of transitions leads to waves with three times the
perturbation frequency (3:1 resonance). \ }%
\label{resonances}%
\end{center}
\end{figure}

\ Figure \ref{resonances} shows several examples of this phenomenon, in which
quite subtle modulations of a stationary wave pattern become magnified with
increasing downstream distance and lead to the breaking and reconnection of
wave fronts in the downstream region. \ The simulations in these examples were
all done at an average flow velocity of $v_{0}=3$. \ By comparison, the
boundary between absolute and convective instability of the Hopf/FDO
instability occurs at $v_{AC}\approx2.82$. \ \ Stationary waves controlled by
the boundary remain possible at velocities well below this threshold,
however.\cite{Kuptsov} \ Thus, while the flow velocity is far from the
kinematic limit, it lies well within the regime where boundary-controlled
stationary waves are stable in the absence of flow modulation. \ The minimum
velocity $v_{0}-\delta_{v}$ never falls below $v_{AC}$ except in figure
\ref{resonances}A, \ and then only by a small amount.\ \ 

In figure \ref{resonances}A, \ the perturbation frequency is very close to the
natural frequency of the chemical oscillator. \ \ The effect of the flow
modulation is visible in this space-time plot as a slight pulsation of the
wavefronts. \ The pulsation becomes stronger at positions farther downstream,
until there is a transition to a region of nearly uniform synchronous
oscillation, synchronized to the period of the flow modulation. \ If the
modulation frequency is changed, \ \ the pattern in the downstream region
remains synchronized to the modulation, and the result is either upstream or
downstream travelling waves. \ An example of upstream waves is shown in figure
\ref{resonances}B. \ In this case the velocity modulation is at a frequency
lower than the intrinsic natural frequency of the medium. \ \ As one can see
from the figure, the system's response to the velocity perturbation is
nonlinear. \ Instead of a simple sinusoidal displacement, the stationary
wavefronts develop a series of sharp cusps. \ At a certain downstream
position, the wavefronts break and reconnect, and the periodic disturbances
become the source of a set of travelling waves with frequency equal to the
modulation frequency, just as if the boundary were being driven at that
frequency. \ 

Just as in the case of ordinary FDO phase waves driven by a perturbation at
the boundary \cite{Kaern1}\cite{Kaern2}\cite{Kaern5}, \ perturbations slower
than the intrinsic frequency give rise to upstream travelling waves. \ (In
general, the phase velocity, wavelength and frequency of the travelling waves
obey the kinematic relationships discussed in \cite{Kaern3}\cite{Kaern4}%
\cite{Faraday}\cite{Kaern5} and \cite{McGraw}.) \ These waves propagate
irregularly in the region just downstream from the transition, \ but with
increasing downstream distance they become smoother. \ \ The temporal
frequency of the waves is locked to the velocity modulation frequency.
\ \ Figure \ref{resonances}C \ shows a more complicated situation with two
consecutive transitions. \ The first transition to travelling waves occurs
much as in \ref{resonances}B. \ However, \ instead of smoothing out with
downstream distance, \ these waves propagate irregularly and develop a second
instability at a position farther downstream, \ leading to travelling waves
with a temporal frequency exactly \emph{twice} that of the velocity
perturbation. \ This can be viewed as a form of $2:1$ frequency locking. \ The
latter travelling wave smooths out with downstream distance and appears to be
the final asymptotic waveform. \ \ An asymptotic waveform at three times the
velocity perturbation frequency is also possible, as in figure
\ref{resonances}D. \ \ In general, \ a sequence of transitions leads to
successive regions of stationary, 1:1, 2:1, 3:1, etc. \ waves. \ In the
particular case of fig. \ref{resonances}D, the three transitions are quite
close together. \ \ Which asymptotic waveform is selected, and the exact
distances from one transition to the next, \ depend in nontrivial ways on the
perturbation frequency and amplitude. \ We will explore this dependence in
subsequent work; \ \ a behavior somewhat analogous to Arnold tongues seems to
occur. \ \ We have observed asymptotic waves in 1:1, 2:1 and 3:1 ratios to the
perturbation frequency, \ but we have not yet observed other rational
multiples such as 2/3. \ \ Interestingly, the tendency of stationary waves to
break is strongest not at the intrinsic natural frequency of the oscillator,
but at approximately $0.65\omega_{0}$.\ 

Immediately downstream from any transition point, the travelling waves
generally propagate with a pulsating phase velocity, but become smoother with
increasing distance downstream. \ Such behavior was observed in both
experiments and numerical simulations for waves forced at the boundary under a
steady flow velocity \cite{Kaern2}. \ In that case, the pulsating phase
velocity was due to a mismatch between the oscillations driving the waves and
the limit cycle of the intrinsic dynamics in the flow reactor. \ The
explanation in this case is the same. \ Instead of being driven by an
oscillation at the inflow boundary, however, these travelling waves are driven
by an oscillation \emph{induced} by the flow velocity modulation. \ As in the
case without flow modulation, \ diffusion tends to smooth the jumping waves as
they travel downstream, \ unless the velocity perturbation induces a second
instability as in figures \ref{resonances}C,D. \ 

The spatiotemporal resonance manifested in wave-front disruption and frequency
locking is novel. While previous studies of spatiotemporal resonance involved
direct, global perturbations of the local dynamics \cite{STR1}\cite{STR2}, in
the present case the perturbation acts \textit{only} at the inflow boundary.
This becomes more evident when one considers the equivalent growing
reaction-diffusion system in the co-moving frame \cite{Santiago}%
\cite{Faraday}, where the velocity does not enter into the dynamical equations
except via the boundary condition. Yet this boundary effect propagates into
the spatial domain where it leads the breakup of waves and resonant frequency locking.

Due to the above-mentioned equivalence of flow and growth \cite{Kaern4}%
\cite{Faraday} \ the same phenomenon should be observable in experiments such
as those of \cite{Santiago}\cite{Kaern6} \ which use a stationary medium with
a moving boundary, if the velocity of the boundary is modulated. \ \ It may
also be relevant to biological situations \cite{Pulsatile1}-\cite{Pulsatile3}
in which growth is pulsatile. \

\end{document}